\documentclass[reprint,aps,prb,amsmath,amssymb,longbibliography, notitlepage]{revtex4-2}
\usepackage{graphicx}
\usepackage{xcolor}
\usepackage{bm}
\usepackage{soul}
\usepackage[normalem]{ulem}
\usepackage[colorlinks, linkcolor= blue, citecolor = blue, urlcolor=blue]{hyperref}
\usepackage{comment}
\usepackage{physics}
\usepackage{pifont}
\usepackage{multirow}
\usepackage{cancel}
\usepackage{wrapfig}
\usepackage[stable]{footmisc}
\usepackage{array}
\usepackage{hhline}
\usepackage{booktabs}
\usepackage{chemformula}
\usepackage{placeins}
\newcommand{\cmark}{\ding{51}}%
\newcommand{\xmark}{\ding{55}}%
\def\be{\begin{equation}}
\def\ee{\end{equation}}
\def \bea{\begin{eqnarray}}
\def \eea{\end{eqnarray}}
\def \nn{\nonumber}
\def \cal{\mathcal}

\begin{document}

\title{Orbital-Splitter Current in Altermagnets}

\author{Koushik Ghorai}
\email{koushikgh20@iitk.ac.in}
\author{Sayan Sarkar}
\email{sayans21@iitk.ac.in}
\author{Amit Agarwal}
\email{amitag@iitk.ac.in}
\affiliation{Department of Physics, Indian Institute of Technology Kanpur, Kanpur-208016, India}

\begin{abstract}

In collinear altermagnets, the real-space rotational symmetry of opposite spin sublattices generates a large nonrelativistic spin-splitter current. Orbital transport in this setting has remained largely unexplored. Here, we introduce the orbital-splitter current (OSC), an orbital analogue of the spin-splitter current, and derive its Drude and orbital Berry curvature contributions using a density-matrix framework. We show that the $d$-wave altermagnet \ch{FeSb2} realizes a purely intrinsic OSC because mirror symmetries suppress the Drude channel by forcing the orbital magnetic moment to vanish. The OSC response is strongly anisotropic and, for selected field orientations, exceeds the spin-splitter current by nearly a factor of four. We further show that the OSC generates a damping-like torque in an altermagnet-ferromagnet heterostructure and, when combined with the spin-splitter current, significantly reduces the magnetization switching time.

\end{abstract}

\maketitle

\twocolumngrid


\section{Introduction and Motivation}

Altermagnets~\cite{Smejkal_22_prx2, Yugui_24_advFunMat, Sinova_24_natcomm, Fengpan_25_natrev, Nirmalya_26} have attracted considerable attention as a platform for next-generation ultrafast spintronic devices. In contrast to conventional antiferromagnets, where opposite spin sublattices are related by inversion or translation symmetry, in altermagnets, they are connected via real-space rotational symmetries. This symmetry leads to a nonrelativistic, momentum-dependent spin splitting and enables the generation of a pure transverse spin current along specific crystallographic directions. Such efficient charge-to-spin conversion, known as the spin-splitter effect~\cite{Zelezny_21_prl, Smejkal_22_natel, Sayan_ExSSC_26, Sankar_prb_26}, together with large exchange-driven THz spin dynamics and the absence of stray fields due to compensated magnetic order, makes altermagnets particularly suitable for spintronic applications~\cite{zhang_25_natcomm, Han_25_pra, Smejkal_22_prx, Sayan_AM_switching}. Furthermore, the spin splitting can be tuned or switched using external stimuli such as strain~\cite{Zhang_25_prb} and electric fields~\cite{LeiWang_25_nanoLett, Qihang_25_prl, Yugui_24_prl}, further broadening the scope of altermagnetic systems.

Despite extensive studies on spin transport and charge responses~\cite{smejkal_20_sciadv, Smejkal_22_natel, Ezawa_24_prb, Vanderbrink_24_prl, Sanjay_25_prb}, the role of orbital degrees of freedom in altermagnets remains largely unexplored. This is notable given that orbital contributions have been shown to be comparable to, and in many cases exceed, their spin counterparts across a variety of charge~\cite{Kamal_21_prb, Shibalik_22_prb, Shengyuan_25_prb, Zesheng_25_prb, Koushik_PHE, Koushik_26_arxiv, Sunit_26_arxiv} and magnetization~\cite{Kontani_08_prl, Kontani_09_prl, Oppeneer_19_natcomm, Annika_21_prr} responses. Unlike spin currents, orbital currents do not rely on spin--orbit coupling, and the associated orbital torque exhibits long-range behaviour~\cite{HyunLee_21_natcomm, Dong_Go_23_prl, DongGo_23_commphys, Arnab_23_prb, Mokrousov_24_natphys, Arnab_25_natcomm, Shengyuan_25_arxiv}. These features motivate a specific question: can altermagnets support a charge-neutral transverse orbital current alongside the spin-splitter current (SSC), and what governs its microscopic origin?

\begin{figure}[t!]
    \centering
    \includegraphics[width=0.8\linewidth]{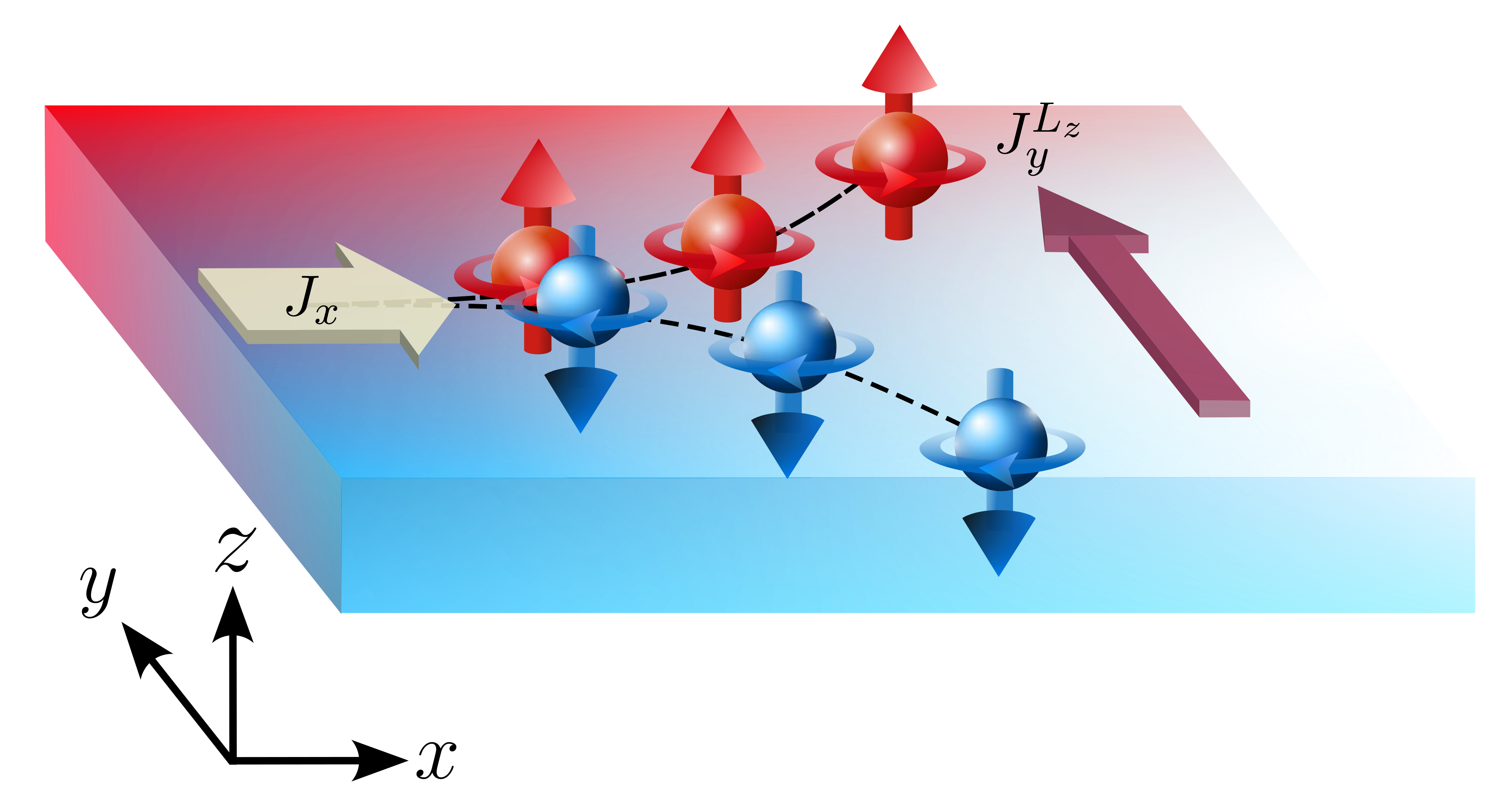}
    \caption{{\bf Schematic of the orbital-splitter effect.} A longitudinal charge current $J_x$ in a $d$-wave altermagnet generates a pure transverse orbital current $J^{L_z}_y$, without any accompanying charge flow. This drives oppositely polarized orbital angular momenta $L_z$ (red and blue arrows) toward the transverse edges of the sample.}
    \label{Fig:Fig_Schematic}
\end{figure}

Here, we answer this question affirmatively by introducing the orbital-splitter current (OSC). The OSC corresponds to a transverse flow of orbital angular momentum (OAM) without any accompanying transverse charge current (see Fig. \ref{Fig:Fig_Schematic}). For selected field orientations, the longitudinal orbital component also vanishes, resulting in a purely transverse response. Microscopically, the OSC originates from two distinct mechanisms: a Drude contribution arising from the electric-field-induced shift of the Fermi surface and an intrinsic component governed by the orbital Berry curvature (OBC), which reflects the underlying band geometry. 
As a representative example, we consider \ch{FeSb_2}~\cite{Smejkal_21_PNAS}. The crystalline symmetries of \ch{FeSb2} allow an OBC-mediated OSC while suppressing the Drude contribution.
In the absence of spin--orbit coupling (SOC), the OSC in this material is comparable in magnitude to its spin counterpart, and weak SOC further enhances the OSC to nearly four times that of the SSC.

We then analyze the angular dependence of the longitudinal and transverse orbital currents and how they influence the orbital-torque response. We show that in contrast to the SSC, which predominantly generates a field-like torque, the OSC produces a damping-like torque. For realistic electric field strengths, the resulting effective magnetic field is comparable to the magnetic anisotropy, providing an efficient mechanism for magnetization control and switching dynamics. 

The remainder of the paper is organized as follows. In Sec.~\ref{sec:Theory}, we develop the density-matrix formalism for the orbital current. In Sec.~\ref{sec:symmetry_analysis}, we analyze the constraints imposed by fundamental and crystallographic point-group symmetries on the orbital response. In Sec.~\ref{sec:material}, we present the OSC in the altermagnet \ch{FeSb_2} and then examine its angular dependence for in-plane electric field orientations. Section~\ref{sec:orbital_torque} discusses magnetization switching in an adjacent ferromagnet driven by the OSC. We conclude and summarize our findings in Sec.~\ref{sec:conclusion}.

\begingroup
\begin{table*}[t]
\centering
\renewcommand{\arraystretch}{1.5}
\setlength{\tabcolsep}{8pt}
\caption{ The symmetry restrictions of the Drude and intrinsic components of the orbital conductivity (OC). The cross (\xmark) and tick (\cmark) marks indicate that the corresponding response tensor is symmetry forbidden and allowed, respectively. For brevity, we denote the OC $\sigma_{b;c}^{L_a}$ by $abc$, where $\{a,b,c\}=\{x,y,z\}$. Here, ${\cal M}_{a}$ and ${\cal C}_{n}^a$ represent mirror and $n$-fold rotation symmetry operations along the $a$ direction.}
\begin{tabular}{c c c c c c c c c c c c}
\hline\hline
OC & $abc$ & $\mathcal{P}$ & $\mathcal{T}$ & $\mathcal{PT}$ 
& $\mathcal{C}^x_{2,4,6}\, , \mathcal{M}_x$ & $\mathcal{C}^y_{2,4,6}\, , \mathcal{M}_y$ & $\mathcal{C}^z_{2,4,6}\, , \mathcal{M}_z$ & $\mathcal{M}_x\mathcal{T}$ & $\mathcal{M}_y\mathcal{T}$
& $\mathcal{M}_z\mathcal{T}$ & $\mathcal{C}^z_4 \mathcal{T}$\\
\hline\hline 
\noalign{\vskip 2pt}

\multirow{3}{*}{$\sigma^{L_a,\mathrm{Dr}}_{b;c}$} & $zxy$, $zyx$ & \cmark & \xmark & \xmark & \cmark & \cmark & \cmark & \xmark & \xmark & \xmark & \cmark \\[3pt]
\cline{2-12}

& $zxx$, $zyy$  & \cmark & \xmark & \xmark & \xmark & \xmark & \cmark & \cmark & \cmark & \xmark & \cmark \\[3pt]
\hline

\multirow{3}{*}{$\sigma^{L_a,\mathrm{OBC}}_{b;c}$} & $zxy$, $zyx$ & \cmark & \cmark & \cmark & \cmark & \cmark & \cmark & \cmark & \cmark & \cmark & \xmark \\[3pt]
\cline{2-12}

& $zxx$, $zyy$ & \cmark & \cmark & \cmark & \xmark & \xmark & \cmark & \xmark & \xmark & \cmark & \xmark\\
\noalign{\vskip 2pt}
\hline\hline
\end{tabular}

\label{Sym_Tab}

\end{table*}
\endgroup



\section{Theory of orbital current} 
\label{sec:Theory}


In altermagnets, the OSC is a symmetry-guided manifestation of the orbital Hall current (OHC)~\cite{Dimi_AdvPhys_24, Dimi_25_prl}. The OHC refers to the transverse flow of orbital-angular-momentum-polarized charge carriers in response to an applied electric field. Recent work used the term orbital-splitter effect for real-space accumulation of orbital angular momentum in driven multiorbital lattices~\cite{Aase_25_prb}. Our focus is complementary: we formulate the bulk orbital-current conductivity in altermagnets, separate the OSC response into Fermi-surface Drude and intrinsic OBC channels, and identify the symmetry conditions under which the transverse orbital current is not accompanied by transverse charge flow. For carriers carrying OAM $L_a$ and moving with velocity $v_b$, the symmetrized orbital current operator is defined as $\hat{j}_b^{a} = \frac{1}{2}\{\hat{L}_a, \hat{v}_b \}$, where $a$ and $b$ denote Cartesian indices.

In this work, we employ the density-matrix formalism~\cite{Debottam_24_prb, Kamal_21_prb} to derive the orbital current. Within this framework, the orbital current is obtained by taking the trace of the orbital current operator weighted by the nonequilibrium density matrix, $J_b^{L_a} = \Tr(\hat{j}_b^{a} \hat{\rho}^E)$. In the presence of a dc electric field $\bm{E}$, the nonequilibrium density matrix obeys the quantum Liouville equation, which to linear order in the field takes the form
\be \label{rho_1}
\frac{\partial \rho_{nm}^{E}}{\partial t} + \left( \frac{1}{\tau} + i\omega_{nm} \right)\rho_{nm}^{E} = \frac{e}{\hbar}\bm{E}\cdot[\mathcal{D}_{\bm{k}}\rho^{(0)}]_{nm}~.
\ee
Here, $\tau$ denotes the relaxation time, and $\omega_{nm} = (\varepsilon_{n\bm k} - \varepsilon_{m\bm k})/\hbar$ is the interband transition frequency between two unperturbed bands of energy $\varepsilon_{n\bm k}$ and $\varepsilon_{m\bm k}$ at the same Bloch momentum $\bm{k}$. The covariant derivative in momentum space is defined as $[\mathcal{D}_{\bm{k}}\rho^{(0)}]_{nm} = \delta_{nm}\partial_{\bm k}f_{n\bm k} + i\bm{\mathcal R}_{nm}f_{nm}$, where $f_{n\bm{k}}$ is the Fermi-Dirac distribution and $f_{nm}= f_{n\bm k}-f_{m\bm k}$ is the equilibrium population inversion. The interband part of the covariant derivative depends on the band-resolved Berry connection $\bm{\mathcal R}_{nm} = \bra{u_{n\bm k}}i\partial_{\bm k}\ket{u_{m\bm k}}$, where $\ket{u_{n\bm k}}$ is the cell-periodic part of the Bloch eigenstates $\ket{\psi_{n\bm k}} = e^{i\bm{k}\cdot \bm{r}}\ket{u_{n\bm k}}$ for an electron in the $n^{\mathrm{th}}$ band. 

The steady-state solution of the density-matrix equation is
\be
 {\rho}_{nm}^{E} = \frac{e}{\hbar}g_{nm} \left[ \delta_{nm} \bm{E}\cdot \partial_{\bm k} f_n^{(eq)} + i (\bm{E} \cdot \bm{\mathcal{R}}_{nm})f_{nm} \right]~,
\ee
with $g_{nm} = [(1/\tau)+i\omega_{nm}]^{-1}$. The first term of the density-matrix element captures the intraband response, while the second term describes the interband coherence of the Bloch electrons.
Multiplying this density-matrix element by the orbital current operator and tracing over all bands yields the orbital current
\bea \label{OH_current}
J_b^{L_a} &=& \sum_{nm,\bm{k}} j_{nm,b}^a \rho_{mn}^E 
~=~ \tau\frac{e}{\hbar} \int_{n, {\bm k}} j_{n,b}^a (\bm{E} \cdot \partial_{\bm k} f_{n\bm k}) \nn \\
&& +  \frac{e}{\hbar} \int_{n, {\bm k}}~ \sum_{m\neq n}g_{mn} \frac{j_{nm,b}^a (\bm{E} \cdot \bm{v}_{mn})}{\omega_{mn}} f_{mn}~.
\eea
Here, we have defined $\int_{n, {\bm k}} \equiv \sum_n \int d^D{\bm k}/{(2\pi)^D}$, with $D$ being the dimension of the system. The matrix element of the orbital current operator $j_{nm,b}^{L_a} = \frac{1}{2}\{\hat{L}_a, \hat{v}_b \}_{nm}$ is related to the OAM, $\bm{L}_{mn} = m_e\sum_{p\neq n}(\bm{v}_{mp} + \delta_{mp}\bm{v}_n)\times \bm{\cal{R}}_{pn}$~\cite{Cong_prb_21, Cong_21_prb2}, and the unperturbed velocity matrix element $\bm{v}_{nm} = (1/\hbar)\bra{u_{n\bm k}}\partial_{\bm k}{\cal H}_0 \ket{u_{m \bm k}}$. The first term of the orbital current is a Fermi-surface contribution, and it vanishes in insulators, whereas the second term, being a Fermi-sea response, survives in both metals and insulators. In the weak-scattering limit $(\omega_{mn}\tau \gg 1)$, $g_{mn}\approx -i/\omega_{mn}$, and the Fermi-sea term reduces to an intrinsic contribution governed by the OBC,
\be
\Omega_{n\bm k;bc}^{L_a} = \frac{2}{\hbar} \sum_{m\neq n} \Im\left[ \frac{ v_{nm}^c j_{mn,b}^{L_a}}{\omega_{mn}^2} \right]~.
\ee
For convenience, we define the linear orbital conductivity (OC) $J_b^{L_a} = \sigma_{b;c}^{L_a} E_c$ and separate it into two parts according to their physical origin 
\be \label{Eq_orbital_cond}
\sigma_{b;c}^{L_a} = \sigma_{b;c}^{L_a,\mathrm{Dr}} + \sigma_{b;c}^{L_a,\mathrm{OBC}}~.
\ee
Apart from their physical origins, the two contributions also satisfy distinct symmetry constraints, which we discuss in detail in Sec.~\ref{sec:symmetry_analysis}. The Drude term, originating from the field-induced Fermi surface shift, is given by
\be
\sigma_{b;c}^{L_a,\mathrm{Dr}} = \frac{e\tau}{\hbar} \int_{n, {\bm k}} j^{L_a}_{n,b} \partial_c f_{n\bm k}~,
\ee
while the OBC-driven intrinsic component is 
\be
\sigma_{b;c}^{L_a,\mathrm{OBC}} = e \int_{n, {\bm k}} f_{n\bm k} \Omega_{n\bm k;bc}^{L_a}~.
\ee
The corresponding expressions for the spin conductivity (SC) are easily obtained by replacing the OAM with spin angular momentum $\bm{S} = (\hbar/2)\bm{\sigma}$ in Eq. (\ref{Eq_orbital_cond}), where $\bm{\sigma}=\{\sigma_x, \sigma_y, \sigma_z\}$ are the Pauli spin matrices. The spin conductivity $\sigma_{b;c}^{S_a}$ also consists of Drude and intrinsic contributions, with the latter governed by the spin Berry curvature~\cite{Shukla_prb2025, Xiang_FOP2023, sinova_prl2004, Xiang_prl2025}. The total spin conductivity can be expressed as
\be
\sigma_{b;c}^{S_a} = \sigma_{b;c}^{S_a,\mathrm{Dr}} + \sigma_{b;c}^{S_a,\mathrm{SBC}}~.
\ee
Depending on the symmetry, one or both channels of the SC and OC may contribute. The role of symmetry is twofold: it determines which microscopic channels can survive and whether the allowed orbital current is accompanied by unwanted longitudinal orbital or transverse charge responses. We discuss these constraints in detail in the next section.



\section{Symmetry analysis} 
\label{sec:symmetry_analysis}


The emergence of the OSC can be understood from symmetry considerations. We identify the symmetry elements that enable a pure orbital Hall response, without any accompanying longitudinal component, for specific orientations of the electric field. We first illustrate this using a simple example of mirror symmetry and then carry out a systematic crystallographic symmetry analysis of the OC to determine the broader set of symmetry conditions that support the OSC.

Consider a system with a mirror symmetry $\mathcal{M}_x$ about the $x$-axis. For an electric field applied along $\bm{x}$, both the field and the transverse orbital current $J_y^{L_z} \sim L_z v_y$, with out-of-plane OAM $L_z$, change sign under $\mathcal{M}_x$. As a result, the corresponding orbital Hall conductivity $\sigma_{y;x}^{L_z}$, defined through $J_y^{L_z} = \sigma_{y;x}^{L_z} E_x$, remains symmetry-allowed. In contrast, the longitudinal component $\sigma_{x;x}^{L_z}$ is prohibited, since $J_x^{L_z}$ remains invariant while the electric field reverses sign under $\cal{M}_x$. A similar symmetry argument in the charge sector shows that $\mathcal{M}_x$ also suppresses transverse charge response in the $xy$ plane (see Appendix~\ref{Appendix_Sec:Charge_current}). Consequently, when the electric field is applied along the $x$-axis, the system supports a purely transverse orbital current with out-of-plane orbital polarization, i.e., the OSC. This reasoning extends to other mirror-symmetric directions, as well.

We now examine the constraints imposed by fundamental symmetries. The finiteness of OHC or OSC requires the invariance of the constitutive relation $J_b^{L_a} = \sigma_{b;c}^{L_a} E_c$ under space-inversion ($\cal{P}$) and time-reversal ($\cal{T}$). Under $\cal{P}$, both the orbital current and the electric field change sign, leaving the conductivity invariant. Accordingly, inversion symmetry does not restrict the OHC and OSC. Time-reversal symmetry acts differently on the two contributions to the OC. Under $\cal{T}$, $J_b^{L_a} \xrightarrow{\cal{T}} J_b^{L_a}$, $\bm{E} \xrightarrow{\cal{T}} \bm{E}$ and $\tau \xrightarrow{\cal{T}} -\tau$, implying that the Drude component $\sigma_{b;c}^{L_a,\mathrm{Dr}}$ is $\cal{T}$-odd, while the intrinsic OBC contribution $\sigma_{b;c}^{L_a,\mathrm{OBC}}$ is $\cal{T}$-even~\cite{Bhowal_20_prb}. Consequently, the Drude term is finite only in magnetic systems, whereas the intrinsic contribution can exist in both magnetic and nonmagnetic materials.

\begin{figure*}[t]
    \centering
    \includegraphics[width=1\linewidth]{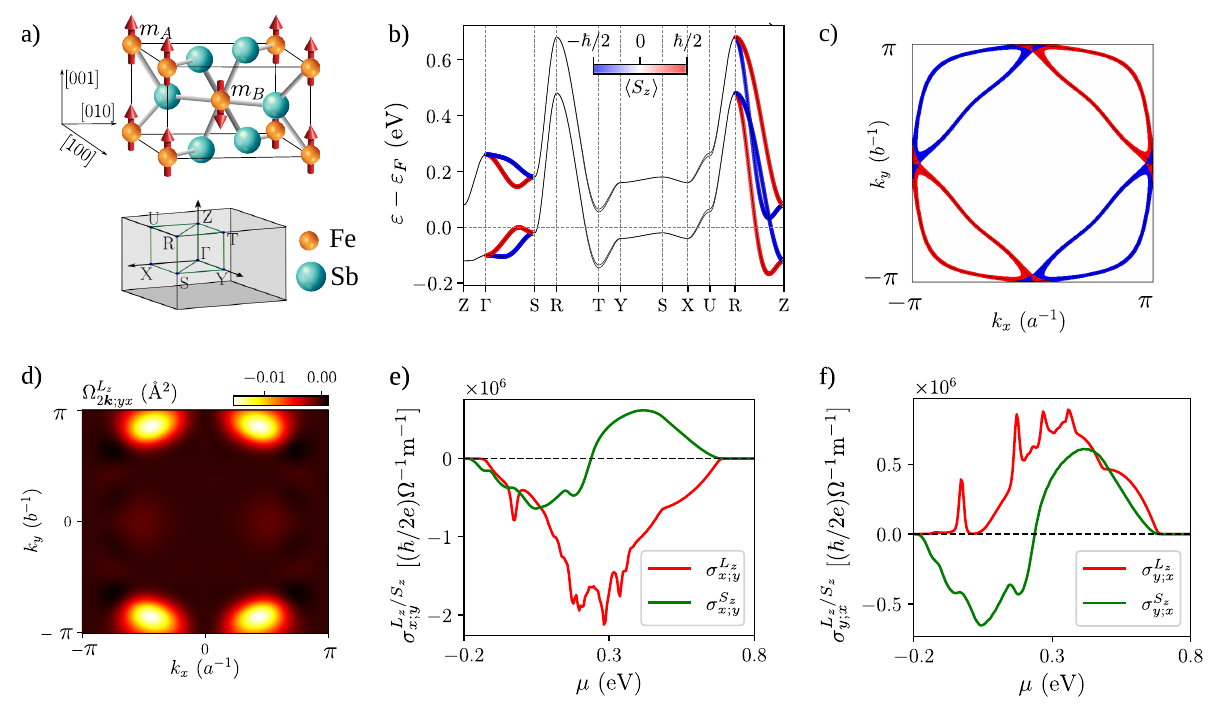}
    \caption{ 
    (a) Crystallographic and magnetic structure of $\mathrm{FeSb_2}$, with red arrows indicating the orientation of magnetic moments on the \ch{Fe} atoms. The three-dimensional Brillouin zone and corresponding high-symmetry points are also shown. (b) Band dispersion highlighting the anisotropic, momentum-dependent spin splitting along the high-symmetry paths $\Gamma$--S and R--Z. (c) Spin-polarized Fermi surface at $\mu = 0.16~\mathrm{eV}$ in the projected $xy$ plane of the Brillouin zone, exhibiting the characteristic $d$-wave splitting. (d) Momentum-space distribution of the orbital Berry curvature $\Omega_{n\bm k;bc}^{L_z}$ for the top valence band. (e), (f) Chemical-potential dependence of the Hall conductivity components $\sigma^{L_z/S_z}_{x;y}$ and $\sigma^{L_z/S_z}_{y;x}$, with orbital and spin contributions shown by red and green curves, respectively. Calculations are performed at $T = 50~\mathrm{K}$ with a scattering time $\tau = 0.1~\mathrm{ps}$.
    }
    \label{Fig1_FeSb2}
\end{figure*}

To analyze the role of crystalline symmetries, we invoke Neumann's principle~\cite{nye1957,vainshtein2013,Gallego_2019}. It states that if a crystal is invariant under a set of symmetry operations, its physical response tensors must also remain invariant under those operations. Both components of the OC are axial in nature, and only $\cal{T}$-symmetry differentiates them. Under a general symmetry operation $\cal{O}$, they transform differently as follows~\cite{newnham2004,Briss_ROP1963,zhang_NSR2023,Brandmuller_CMA1986,Authier_book2013,Litvin_mpg2013,Gallego_2019} 
\bea
\sigma^{L_{a'},\mathrm{Dr}}_{b';c'} &=& \eta_{\mathcal{T}}\det{\mathcal O}{\mathcal O}_{a'a}{\mathcal O}_{b'b}{\mathcal O}_{c'c}~\sigma^{L_a,\mathrm{Dr}}_{b;c}~, \\
\sigma^{L_{a'},\mathrm{OBC}}_{b';c'} &=& \det{\mathcal O}{\mathcal O}_{a'a}{\mathcal O}_{b'b}{\mathcal O}_{c'c}~\sigma^{L_a,\mathrm{OBC}}_{b;c}~,
\eea
Here, $\eta_{\cal T}=\pm 1$ is associated with the magnetic point group symmetry transformation: 
$\eta_{\cal T}=-1$ ($\eta_{\cal T}=1$) for a magnetic (nonmagnetic) point-group operation ${\cal O} \equiv {\cal RT}$ (${\cal O} \equiv {\cal R}$), with $\cal R$ being a spatial operation. In Table~\ref{Sym_Tab}, we list the symmetry-allowed independent components of the OC for OAM polarized along the $z$ direction and transport in the $xy$ plane. A complete listing for other polarization directions and transport geometries is provided in Appendix~\ref{Appendix_Sec:Orbital_current_symmetry}.

The symmetry requirements for the OSC are more restrictive than those for the OHC. The OSC can emerge when the transverse orbital current $\sigma^{L_a}_{b;c}$ ($b \neq c$) is symmetry allowed, while both the longitudinal orbital component $\sigma^{L_a}_{b;b}$ and the transverse charge conductivity $\sigma_{b;c}$ are forbidden. Using the symmetry tables for charge currents (Table~\ref{Table:Charge_conductivity} in Appendix~\ref{Appendix_Sec:Charge_current}) and orbital currents (see Appendix~\ref{Appendix_Sec:Orbital_current_symmetry}), one can determine whether the OSC occurs in a given system. For completeness, we list the magnetic point groups that allow a finite out-of-plane-polarized OSC for the three orthogonal transport planes in Appendix~\ref{Appendix_Sec:MPG_OSC}.

\section{Material example: \texorpdfstring{$\mathbf{FeSb_2}$}{FeSb2}} 
\label{sec:material}


In this section, we evaluate the orbital current in the $d$-wave altermagnet \ch{FeSb2}. For spins polarized along the $z$ axis, the system belongs to the centrosymmetric space group $Pnnm$ and, in the absence of SOC, contains three mirror symmetries~\cite{Smejkal_21_PNAS}. 
Notably, the system does not possess $\cal{PT}$ or $\bm{t}\cal{T}$ symmetries that enforce spin degeneracy across the entire Brillouin zone. Instead, the sublattice-connecting symmetries restrict degeneracies to specific regions in momentum space. 
Figure~\ref{Fig1_FeSb2}(a) illustrates the crystal structure and Brillouin zone of \ch{FeSb2}. The lattice parameters are $a = 5.83~\text{\AA}$, $b = 6.54~\text{\AA}$, and $c = 3.18~\text{\AA}$~\cite{Petrovic_prb2005,Zhang_CPB2025}. To describe this altermagnetic system, we adopt a minimal tight-binding model that captures its essential symmetries and electronic structure, using parameters from Ref.~\cite{Roig_prb2024}. The Hamiltonian reads  
\be\label{Ham_AM}
{\cal H} = \varepsilon_{{\bm k}}^0 + t_{x,{\bm k}} \tau_x + t_{z,{\bm k}} \tau_z + \tau_y \vec{\lambda}_{{\bm k}} \cdot \vec{\sigma} + \tau_z \vec{J} \cdot \vec{\sigma}~,
\ee
where $\tau_i$ and $\sigma_i$ are the Pauli matrices in orbital (sublattice) and spin space, respectively. The term $\varepsilon_{\bm k}^0$ represents the sublattice-independent dispersion, while 
$t_{x,\bm k}$ and $t_{z,\bm k}$ describe inter- and intra-sublattice hopping processes. The vector $\vec{\lambda}_{{\bm k}}$ encodes the intrinsic SOC, and $\vec{J}$ denotes the N\'eel order parameter.

The momentum-dependent coefficients entering into the above Hamiltonian take the form
\begin{equation}\label{eq:model_params}
\begin{aligned}
\varepsilon_{\bm k}^0 &= -\mu+
t_{1x}\cos (k_xa) + t_{1y}\cos (k_yb) + t_2\cos (k_zc)
  \\
&\quad+ t_3\cos (k_xa) \cos (k_yb) + t_{4x}\cos (k_xa) \cos (k_zc)
 \\
&\quad + t_{4y}\cos (k_yb) \cos (k_zc)+t_5\cos (k_xa) \cos (k_yb) \cos (k_zc)~,
\end{aligned}
\end{equation}
with the hopping terms,  
\begin{equation}\label{eq:model_params1}
\begin{aligned}
t_{x,\bm k} &= t_8 \cos\frac{k_xa}{2}\cos\frac{k_yb}{2}\cos\frac{k_zc}{2}~, \\
t_{z,\bm k} &= t_6 \sin (k_xa) \sin (k_yb) + t_7 \sin (k_xa) \sin (k_yb) \cos (k_zc)~,
\end{aligned}
\end{equation}
and the SOC terms,
\bea\label{eq:model_params2}
\lambda_{x,\bm k} &=& \lambda_{x0} \sin\frac{k_xa}{2}\cos\frac{k_yb}{2}\sin\frac{k_zc}{2}~, \nn \\
\lambda_{y,\bm k} &=& \lambda_{y0} \cos\frac{k_xa}{2}\sin\frac{k_yb}{2}\sin\frac{k_zc}{2}~,\nn \\
\lambda_{z,\bm k} &=& \lambda_{z0} \cos\frac{k_xa}{2}\cos\frac{k_yb}{2}\cos\frac{k_zc}{2}~.  
\eea  
For $\mathrm{FeSb_2}$, we use the parameter values $t_{1x}=-0.1,t_{1y}=-0.05,t_2=-0.05, t_3=0.06, t_{4x}=0.1,t_{4y}=0.05,t_5=-0.05,t_6=0.05, t_7=-0.1, t_8=0.15$ and $J_z=0.1$ (all in eV). We also assume finite SOC strengths $\lambda_{x0}=\lambda_{y0}=0.01~\mathrm{eV}$ and $\lambda_{z0} = 0.005~\mathrm{eV}$. 
%


\begin{figure*}[t!]
    \centering
    \includegraphics[width=0.85\linewidth]{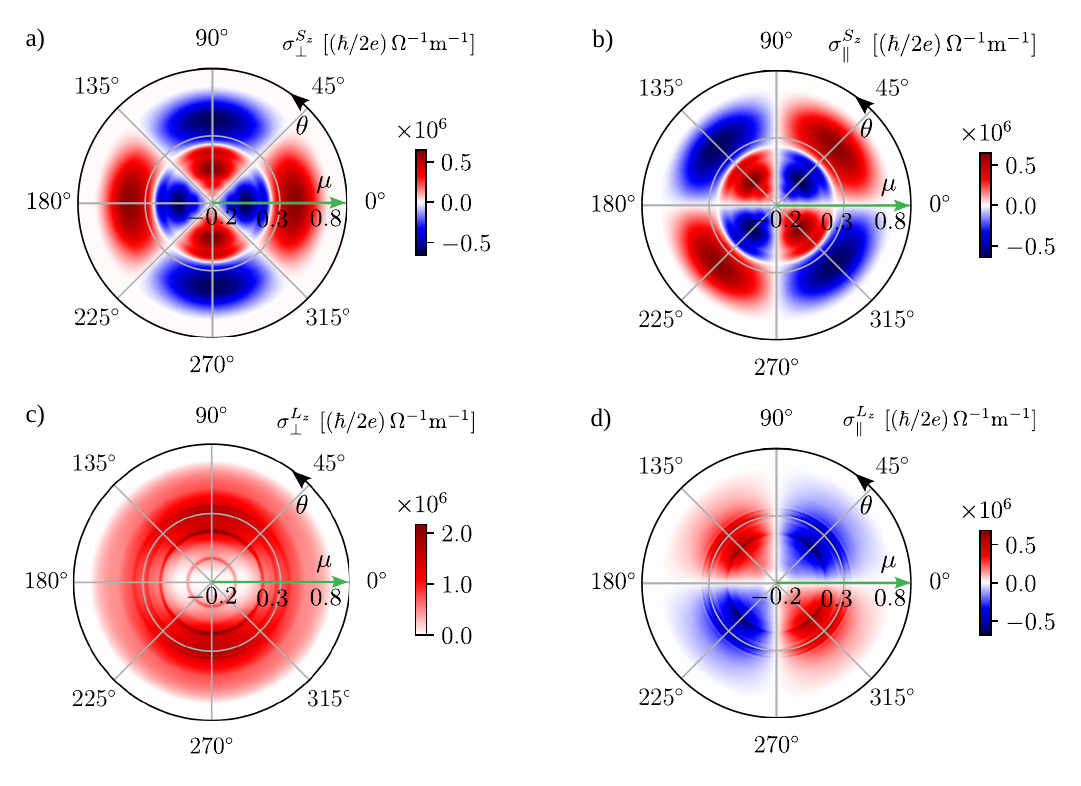}
    \caption{
    (a), (b) Transverse and longitudinal spin currents, flowing perpendicular and parallel to the applied electric field, respectively, as the field is rotated within the $xy$ plane. The radial axis of the polar plots is the chemical potential. (c), (d) Corresponding transverse and longitudinal orbital currents for the same in-plane field orientations. Notably, when the field is aligned along the $\pm x$ and $\pm y$ directions, the longitudinal components of both spin and orbital currents vanish, providing characteristic signatures of the spin-splitter and orbital-splitter currents, respectively. Calculations are performed at $T = 50~\mathrm{K}$ with a scattering time $\tau = 0.1~\mathrm{ps}$.
    }
    \label{Fig2_FeSb2}
\end{figure*}
The energy eigenvalues for the above Hamiltonian are
\bea
\varepsilon_{\alpha = \pm, \beta = \pm} &=& \varepsilon_{\bm k}^0 + \alpha \biggl( t_{x,{\bm k}}^2 + t_{z,{\bm k}}^2 + \vec{\lambda}_{{\bm k}}^2 + \vec{J}^2 \nn \\
&& + 2\beta \sqrt{t_{z,{\bm k}}^2\vec{J}^2 + (\vec{\lambda}_{{\bm k}} \times \vec{J})^2} \biggr)^{1/2}~.
\eea
The band structure, presented in Fig.~\ref{Fig1_FeSb2}(b), exhibits the characteristic altermagnetic spin splitting along the high-symmetry $\Gamma$--S and R--Z paths. This feature is also reflected in the Fermi surface at $\mu = 0.16$ eV, shown in Fig.~\ref{Fig1_FeSb2}(c), which exhibits an even-parity, spin-polarized $d$-wave pattern. This nonrelativistic spin splitting gives rise to a large transverse pure spin current via the Drude mechanism for charge currents applied along the $\bm{x}$ or $\bm{y}$ directions~\cite{Zelezny_21_prl, Smejkal_22_natel}. We consider charge transport in the $xy$ plane with OAM polarized along the $z$ axis. In this configuration, the two independent Hall components are $\sigma^{L_z/S_z}_{x;y}$ and $\sigma^{L_z/S_z}_{y;x}$. In the absence of relativistic SOC, the mirror symmetries $\{ \cal{M}_x, \cal{M}_y, \cal{M}_z \}$ enforce a vanishing orbital moment throughout the Brillouin zone. As a consequence, the Drude contribution to the OC vanishes, and the orbital current is entirely governed by the OBC, whose momentum-space distribution is shown in Fig.~\ref{Fig1_FeSb2}(d). As follows from the symmetry analysis in Sec.~\ref{sec:symmetry_analysis}, these mirror symmetries impose additional constraints. They restrict the OBC-driven longitudinal orbital current for fields applied along the $x$ or $y$ directions and also enforce a vanishing Berry curvature, thereby suppressing the anomalous charge Hall response. These symmetries also forbid the Drude charge Hall current (see Appendix~\ref{Appendix_Sec:Orbital_charge_current_NoSOC}). Consequently, \ch{FeSb2} supports a purely intrinsic transverse orbital current, without accompanying transverse charge or longitudinal orbital components. This is the orbital analogue of the SSC, namely the OSC in \ch{FeSb2}.

Weak SOC lifts some degeneracies along high-symmetry planes, lines, and points and also induces a finite orbital moment. However, the corresponding Drude contribution to the orbital current remains negligible compared with the OBC-driven part. The induced charge current generated by the broken mirror symmetries is likewise small (see Appendix~\ref{Appendix_Sec:Charge_current_FeSb2}). At the same time, the SOC-induced gaps enhance the orbital response because the OBC scales inversely with the band gap. The chemical-potential dependence of the orbital and spin contributions to the two Hall conductivity components, $\sigma^{L_z/S_z}_{x;y}$ and $\sigma^{L_z/S_z}_{y;x}$, is shown in Figs.~\ref{Fig1_FeSb2}(e) and \ref{Fig1_FeSb2}(f). While the orbital and spin responses are comparable in magnitude for $\sigma^{L_z/S_z}_{y;x}$, the orbital contribution to $\sigma^{L_z/S_z}_{x;y}$ is substantially larger, reaching nearly $3.4$ times the corresponding spin current, with a maximum value of $-2.16 \times 10^6~(\hbar/2e)\Omega^{-1}\mathrm{m}^{-1}$. The two responses also show distinct chemical-potential dependence, reflecting the anisotropy of the material.

\subsection{Angular dependence of orbital and spin currents} \label{sec:angular_dependence}

The anisotropy of the system can be probed by examining the orbital and spin currents for different in-plane orientations of the electric field. We illustrate this for transport in the $xy$ plane; the same construction applies to other transport planes. The orbital current can be split into a transverse component, perpendicular to the applied field, and a longitudinal component, parallel to it. The transverse component, defined as $\bm{J}_{\perp}^{X_a} = \sigma_{\perp}^{X_a}\bm{E}$, is given by
\be \label{sigma_perp}
\sigma^{X_a}_{\perp} = - \sigma^{X_a}_{x;y}\sin^2\theta + \sigma^{X_a}_{y;x}\cos^2\theta - \frac{1}{2}(\sigma^{X_a}_{x;x} - \sigma^{X_a}_{y;y}) \sin2\theta~,
\ee
where $\theta$ is the angle made by the electric field with the $x$ axis, and $X_a = L_a$ ($S_a$) denotes orbital (spin) angular momentum. The longitudinal component, defined along the electric field direction as $\bm{J}_{\parallel}^{X_a} = \sigma_{\parallel}^{X_a}\bm{E}$, takes the form
\be \label{sigma_parallel}
\sigma^{X_a}_{\parallel} = \sigma^{X_a}_{x;x}\cos^2\theta + \sigma^{X_a}_{y;y} \sin^2\theta + \frac{1}{2}(\sigma^{X_a}_{x;y} + \sigma^{X_a}_{y;x}) \sin 2\theta~.
\ee
Figures~\ref{Fig2_FeSb2}(a) and \ref{Fig2_FeSb2}(b) show the angular dependence of the transverse and longitudinal spin currents. As the electric field rotates, the spin current follows the alternating spin polarization of the Fermi surface and reverses sign between successive lobes. For $\theta = 0, \pi/2, \pi$, and $3\pi/2$, the transverse spin current is maximal while the longitudinal component vanishes, providing a clear signature of the SSC. The orbital response evolves differently. The transverse orbital current in Fig.~\ref{Fig2_FeSb2}(c) retains the same sign for all field orientations, whereas the longitudinal component in Fig.~\ref{Fig2_FeSb2}(d) shows a four-lobed structure with alternating sign. The latter vanishes at the same symmetry-related angles, reflecting the characteristic behavior of the OSC in \ch{FeSb2}. Both spin and orbital currents are therefore $\pi$ periodic, consistent with the $\sin 2\theta$ and $\cos 2\theta$ terms in Eqs.~(\ref{sigma_perp}) and (\ref{sigma_parallel}).

We further examine the dependence of the transverse spin and orbital conductivities on the N\'eel order parameter $J_z$. As shown in Fig.~\ref{Fig_Orbital_Torque}(a), the spin Hall response varies linearly with the magnetic order and reverses sign upon inversion of the N\'eel vector, consistent with its $\cal{T}$-odd character. By contrast, the orbital Hall response, dominated by the intrinsic OBC mechanism, remains unchanged under $J_z$ reversal. Correspondingly, the orbital Hall current is approximately quadratic in the magnetic order for small $J_z$. Such a response is known to generate damping-like torques~\cite{Farokhnezhad_IOP2023, Oppeneer_26_jap}.


\section{OSC-induced magnetization switching} 
\label{sec:orbital_torque}

\begin{figure*}[t!]
\centering
\includegraphics[width=1\linewidth]{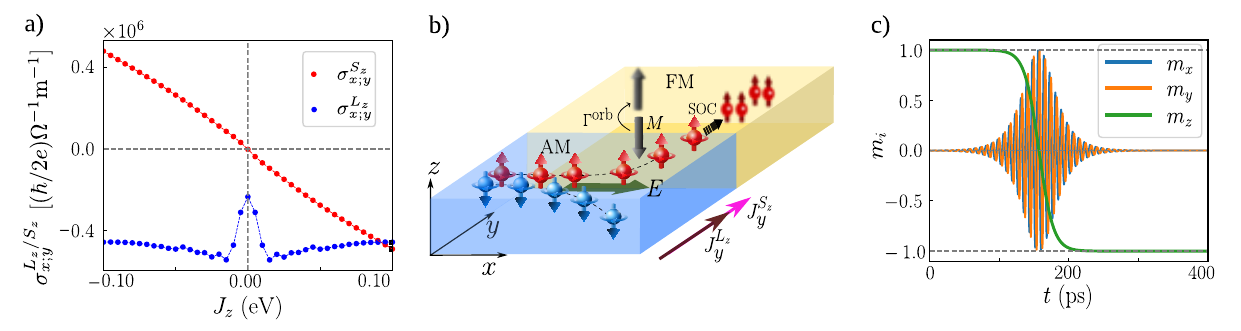}
\caption{(a) Variation of spin and orbital Hall conductivities with the altermagnetic order strength $J_z$. The orbital response does not change sign under reversal of $J_z$, confirming the dominant $\cal T$-even contribution in the system. (b) Schematic of an altermagnet (AM)--ferromagnet (FM) heterostructure, where the AM produces a transverse spin current as well as an accompanying transverse orbital current. Within the FM, spin--orbit coupling transforms part of this orbital current into spin current, contributing to spin torques that switch the ferromagnetic magnetization ($\bm{M}$). (c) Time evolution of the FGT magnetization components under the combined spin-splitter and orbital-splitter current induced torques generated in \ch{FeSb2}.}
\label{Fig_Orbital_Torque}
\end{figure*}

The generation and manipulation of orbital currents have recently emerged as a route to current-induced magnetization switching. In an altermagnet--ferromagnet (AM--FM) heterostructure, as shown in Fig.~\ref{Fig_Orbital_Torque}(b), the orbital current generated in the AM is injected into the adjacent FM, where SOC converts orbital angular momentum into spin angular momentum, producing an effective torque on the magnetization~\cite{HyunLee_21_natcomm, Yang_NC2024, Zhang_NC2025, Teng_TIM2025}. This mechanism remains efficient even when the orbital-current source has weak intrinsic SOC~\cite{Lee_CP2021, Yang_NC2024}.

In the present work, \ch{FeSb2} serves as the orbital-current source. As shown in Fig.~\ref{Fig1_FeSb2}(e), the orbital Hall conductivity reaches values as large as $\sigma^z_{x;y}\sim -2\times 10^6~(\hbar/2e)\Omega^{-1}\mathrm{m}^{-1}$ at $\mu\sim 0.2~\mathrm{eV}$. For an applied electric field $\bm{E}=10^6~\mathrm{V/m}\,\hat{y}$, this corresponds to a transverse orbital current density $J^{L_z}_x\sim -2\times10^{12}~\hbar/2e~\mathrm{A/m^2}$. Upon injection into an adjacent ferromagnetic layer such as \ch{Fe3GaTe2} (FGT), a fraction of this orbital current is converted into a spin current via SOC. Using a conversion efficiency of $\eta\sim 38\%$ in FGT~\cite{X_C_Xie_26_arxiv}, we estimate an effective spin current density of $J^{S_z,\mathrm{orb}}_x\sim 7.6\times 10^{11}~\hbar/2e~\mathrm{A/m^2}$. This adds to the spin current injected through the SSC into the FM, $J^{S_z,\mathrm{spin}}_x\sim 5.0\times 10^{11}~\hbar/2e~\mathrm{A/m^2}$, for the same applied field. These spin currents generate spin torques large enough to switch the FM magnetization.

To explicitly demonstrate this mechanism, we simulate the magnetization dynamics using the Landau--Lifshitz--Gilbert (LLG) equation~\cite{slonczewski_JOM1996, Liu_prl2011, Yuan_IOP2020, Xu_JAP2023}:
\be \label{Dynamic_FM}
\frac{d \hat{m}}{dt} = -\gamma (\hat{m}\times {\bm B}_{\mathrm{eff}}) + \alpha \left(\hat{m} \times \frac{d \hat{m}}{dt}\right) + {\bm \tau}^{\mathrm{FL}} + {\bm \tau}^{\mathrm{DL}},
\ee
where $\hat{m}$ is the unit vector along the magnetization of the FM. $\alpha$ denotes the Gilbert damping parameter. We use $\alpha=0.007$ for \ch{FGT}~\cite{Wang_prb2025}. The effective magnetic field ${\bm B}_{\mathrm{eff}} = {\bm B}_{\mathrm{ans}} + {\bm B}_{\mathrm{ext}}$ includes both anisotropy and external contributions. For FGT, the out-of-plane anisotropy field is $B_{\mathrm{ans}}\sim 3$--$4~\mathrm{T}$~\cite{Zhang_NC2022}. The field-like (FL) and damping-like (DL) torque components are given by
${\bm \tau}^{\mathrm{FL}} = |\Gamma| \xi^{\mathrm{FL}} (\hat{m} \times \hat{\sigma})$ and
${\bm \tau}^{\mathrm{DL}} = |\Gamma| \xi^{\mathrm{DL}}\, \hat{m} \times (\hat{m} \times \hat{\sigma})$, where $\hat{\sigma}$ denotes the spin polarization direction and $\xi^{\mathrm{FL/DL}}$ are the corresponding efficiencies. We define $|\Gamma|=\gamma B^s_{\mathrm{eff}}$, with the effective spin-torque field
\be
B^s_{\mathrm{eff}}=\frac{(\hbar/2e)|J^{S_z}_x|}{M_s l}~,
\ee
where $J^{S_z}_x$ is the total injected spin-current density in units of $(\hbar/2e)\mathrm{A/m^2}$, $M_s$ is the saturation magnetization of the FM, and $l$ is the FM thickness~\cite{Liu_prl2011,Sinova_rmp15}. Using $J^{S_z}_x\simeq 1.26\times10^{12}~\mathrm{A/m^2}$ for the combined SSC and OSC channels, $l\sim 5~\mathrm{nm}$, and a representative FGT value $M_s\sim 3\times10^5~\mathrm{A/m}$~\cite{Zhang_NC2022,Wang_NC2025}, we obtain $B^s_{\mathrm{eff}}\sim 0.3~\mathrm{T}$. In our simulations, we take $\xi^{\mathrm{DL}}/\xi^{\mathrm{FL}}=1$. With these parameters, the magnetization exhibits a clear $180^\circ$ switching within about $200~\mathrm{ps}$, as shown in Fig.~\ref{Fig_Orbital_Torque}(c), compared with about $550~\mathrm{ps}$ for the SSC-only case (see Appendix~\ref{Appendix_Sec:M_switching}). The switching dynamics are primarily governed by the damping-like torque, which counteracts the magnetic anisotropy field. Accordingly, efficient switching requires the condition $B^s_{\mathrm{eff}} > \alpha B_{\mathrm{ans}}$, which is satisfied in the present case.


\section{Conclusion}  
\label{sec:conclusion}


In conclusion, we have introduced the orbital-splitter current as a charge-neutral transverse flow of orbital angular momentum, with two microscopic contributions: an extrinsic Fermi surface shift term and an intrinsic orbital Berry curvature term. Symmetry analysis shows that altermagnets naturally satisfy the conditions for this response, and $\ch{FeSb_2}$ provides a concrete example in which mirror symmetries suppress the Drude channel and leave a purely intrinsic orbital-splitter current. The response is anisotropic, can exceed the spin counterpart for selected field orientations, and can drive damping-like torques in an adjacent ferromagnet. These results identify altermagnets as a promising platform for orbital-current transport and switching.


\bigskip 

\section{Acknowledgments}

K.G. acknowledges the Ministry of Education, Government of India, for financial support through the Prime Minister's Research Fellowship. S.S. is supported by the Indian Institute of Technology Kanpur. A.A. acknowledges funding from the Core Research Grant by ANRF (Sanction No. CRG/2023/007003), Department of Science and Technology, India.




\onecolumngrid

\appendix


\section{Linear charge current and its crystallographic symmetry} \label{Appendix_Sec:Charge_current}


In this section, we present the crystallographic symmetry of the linear charge current. We begin with a brief discussion of its underlying mechanisms. Within the density-matrix formalism, the linear charge current is expressed as
\bea \label{Charge_current}
J_b &=& -e\sum_{nm,\bm{k}} v_{nm}^b \rho_{mn}^E 
= -\tau\frac{e^2}{\hbar} \int_{n, {\bm k}} v_n^b (\bm{E} \cdot \partial_{\bm k} f_{n\bm k}) - \frac{ie^2}{\hbar} \int_{n, {\bm k}}~ \sum_{m\neq n} \frac{v_{nm}^b (\bm{E} \cdot \bm{v}_{mn})}{\omega_{mn}^2} f_{nm}~.
\eea
The quantum Liouville equation and its first-order (in $E$) solution $\rho_{mn}^E$ were introduced in Sec.~\ref{sec:Theory}. The two contributions in Eq.~(\ref{Charge_current}) originate from distinct physical mechanisms. The first term, which depends on the scattering time $\tau$, arises from the field-induced shift of the Fermi surface. The second term is scattering-independent and stems from the Berry curvature of the occupied states~\cite{Mandar_24_nrm}. For convenience, we express the total charge conductivity $J_b = \sigma_{b;c} E_c$ as $\sigma_{b;c} = \sigma_{b;c}^{\mathrm{Dr}} + \sigma_{b;c}^{\mathrm{BC}}$, with
\bea
\sigma_{b;c}^{\mathrm{Dr}} &=& -\tau\frac{e^2}{\hbar} \int_{n, {\bm k}} v_n^b (\partial_c f_{n\bm k})~, \label{sigma_Dr} \\
\sigma_{b;c}^{\mathrm{BC}} &=& - \frac{e^2}{\hbar} \int_{n, {\bm k}}~\Omega_{n}^{bc} f_n~,\label{sigma_BC}
\eea
where the Berry curvature is given by
\be
\Omega_{n}^{bc} = -2\mathrm{Im} \left( \sum_{m\neq n} \frac{v_{nm}^b v_{mn}^c}{\omega_{nm}^2} \right)~.
\ee
The two contributions differ not only in their physical origin but also in their symmetry properties. The Drude term is even under time-reversal symmetry ($\cal{T}$), whereas the Berry-curvature contribution is $\cal{T}$-odd and vanishes in nonmagnetic systems. Consequently, $\sigma_{b;c}^{\mathrm{Dr}}$ transforms identically under both nonmagnetic ($\cal{R}$) and magnetic ($\cal{RT}$) point-group operations, with $\cal R$ denoting a spatial symmetry. In contrast, $\sigma_{b;c}^{\mathrm{BC}}$ obeys distinct symmetry constraints for nonmagnetic and magnetic point groups. The symmetry constraints of these second-rank polar tensors under a general crystallographic point-group operation $\cal{O}$ follow from the transformation rules
\bea \sigma^{\mathrm{Dr}}_{b';c'} = {\mathcal O}_{b'b}{\mathcal O}_{c'c}~\sigma^{\mathrm{Dr}}_{b;c}~, \quad \text{and} \quad 
\sigma^{\mathrm{BC}}_{b';c'} = \eta_{\mathcal{T}}{\mathcal O}_{b'b}{\mathcal O}_{c'c}~\sigma^{\mathrm{BC}}_{b;c}~. \eea 
where $\eta_{\cal T}=\pm 1$ encodes the time-reversal character of the operation: $\eta_{\cal T}=1$ for nonmagnetic and $\eta_{\cal T}=-1$ for magnetic point-group operations. The symmetry properties of the two conductivity channels are summarized in Table~\ref{Table:Charge_conductivity}.


\begin{table}[h]
\centering
\renewcommand{\arraystretch}{1.4}
\setlength{\tabcolsep}{5pt}
\caption{ The symmetry constraints on the two linear charge conductivity components (CC), namely the Drude and Berry-curvature-driven intrinsic contributions. The cross (\xmark) and tick (\cmark) marks indicate that the corresponding response tensor is symmetry forbidden and allowed, respectively. For brevity, we denote the electrical conductivity $\sigma_{b;c}$ by $bc$, where $\{b,c\}=\{x,y,z\}$. Here, ${\cal M}_{b}$ and ${\cal C}_{n}^b$ represent mirror and $n$-fold rotation symmetry operations along the $b$ direction.} 
\begin{tabular}{c c c c c c c c c c c c c c c c c c c c c}
\hline\hline
$\mathrm{CC}$ & $\sigma_{b;c}$ & $\mathcal{P}$ & $\mathcal{T}$ & $\mathcal{PT}$ 
& $\mathcal{C}^x_2,\mathcal{M}_x$ & $\mathcal{C}^y_2,\mathcal{M}_y$ & $\mathcal{C}^z_2,\mathcal{M}_z$ & $\cal{C}_4^x$ & $\cal{C}_4^y$ & $\cal{C}_4^z$ & $\cal{C}_6^x$ & $\cal{C}_6^y$ & $\cal{C}_6^z$ & $\mathcal{M}_x\mathcal{T}$ & $\mathcal{M}_y\mathcal{T}$
& $\mathcal{M}_z\mathcal{T}$ & $\mathcal{C}^z_4 \mathcal{T}$ \\ 
\hline\hline 
\noalign{\vskip 2pt}
\multirow{4}{*}{$\sigma^{\mathrm{Dr}}_{b;c}$}
& $xy$, $yx$ & \cmark & \cmark & \cmark & \xmark & \xmark & \cmark & \xmark & \xmark & \xmark & \xmark & \xmark & \xmark & \xmark & \xmark & \cmark & \xmark \\[3pt]
\cline{2-18}

& $xx$, $yy$ &  \cmark & \cmark & \cmark & \cmark & \cmark & \cmark & \cmark & \cmark & \cmark & \cmark & \cmark & \cmark & \cmark & \cmark & \cmark & \cmark \\[3pt]
\cline{2-18}

& $zx,~xz$ & \cmark & \cmark & \cmark & \xmark & \cmark & \xmark & \xmark & \xmark & \xmark & \xmark & \xmark & \xmark & \xmark & \cmark & \xmark & \xmark \\[3pt]
\cline{2-18}

& $zy,~yz$ & \cmark & \cmark & \cmark & \cmark & \xmark & \xmark & \xmark & \xmark & \xmark & \xmark & \xmark & \xmark & \cmark & \xmark & \xmark & \xmark \\[3pt]
\cline{2-18}

& $zz$ & \cmark & \cmark & \cmark & \cmark & \cmark & \cmark & \cmark & \cmark & \cmark & \cmark & \cmark & \cmark & \cmark & \cmark & \cmark & \cmark \\
\hline

\multirow{4}{*}{$\sigma^{\mathrm{BC}}_{b;c}$}
& $xy$, $yx$ & \cmark & \xmark & \xmark & \xmark & \xmark & \cmark & \xmark & \xmark & \cmark & \xmark & \xmark & \cmark & \cmark & \cmark & \xmark & \xmark \\[3pt]
\cline{2-18}

& $xx$, $yy$ &  \cmark & \xmark & \xmark & \xmark & \xmark & \xmark & \xmark & \xmark & \xmark & \xmark & \xmark & \xmark & \xmark & \xmark & \xmark & \xmark \\[3pt]
\cline{2-18}

& $zx,xz$ & \cmark & \xmark & \xmark & \xmark & \cmark & \xmark & \xmark & \cmark & \xmark & \xmark & \cmark & \xmark & \cmark & \xmark & \cmark & \xmark \\[3pt]
\cline{2-18}

& $zy, yz$ & \cmark & \xmark & \xmark & \cmark & \xmark & \xmark & \cmark & \xmark & \xmark & \cmark & \xmark & \xmark & \xmark & \cmark & \cmark & \xmark \\[3pt]
\cline{2-18}

& $zz$ & \cmark & \xmark & \xmark & \xmark & \xmark & \xmark & \xmark & \xmark & \xmark & \xmark & \xmark & \xmark & \xmark & \xmark & \xmark & \xmark \\
\hline\hline
\end{tabular}

\label{Table:Charge_conductivity}
\end{table}

\newpage


\section{Crystallographic symmetry of all orbital conductivity components} \label{Appendix_Sec:Orbital_current_symmetry}


The tensorial character of the orbital conductivity differs from that of the charge conductivity. The charge current in a system is given by $\langle \hat{j}_a \rangle = \langle -e \hat{v}_a \rangle = \sigma_{a;b}E_b$, whereas the orbital current is defined as $\langle \hat{j}_b^{a}\rangle = \left\langle \frac{1}{2}\{\hat{L}_a, \hat{v}_b \} \right\rangle = \sigma_{b;c}^{L_a} E_c$. Here, $\hat{v}_a$ and $\hat{L}_a$ correspond to the velocity and OAM operators, respectively, and $\{a,b,c\}$ denote Cartesian indices. Because the constitutive relation contains an odd number of axial quantities, both the intrinsic and extrinsic contributions to the orbital conductivity are axial in nature. Furthermore, due to the $\cal{T}$-odd character of OAM, the Drude contribution is $\cal{T}$-odd, while the OBC-driven contribution is $\cal{T}$-even. Nonzero orbital conductivity components can be obtained by solving the transformation rules
\bea
\sigma^{L_{a'},\mathrm{Dr}}_{b';c'} &=& \eta_{\mathcal{T}}\det{\mathcal O}{\mathcal O}_{a'a}{\mathcal O}_{b'b}{\mathcal O}_{c'c}~\sigma^{L_a,\mathrm{Dr}}_{b;c}~, \\
\sigma^{L_{a'},\mathrm{OBC}}_{b';c'} &=& \det{\mathcal O}{\mathcal O}_{a'a}{\mathcal O}_{b'b}{\mathcal O}_{c'c}~\sigma^{L_a,\mathrm{OBC}}_{b;c}~,
\eea
Here, $\eta_{\cal T}=\pm 1$ is associated with the magnetic point group symmetry transformation: 
$\eta_{\cal T}=-1$ ($\eta_{\cal T}=1$) for a magnetic (nonmagnetic) point-group operation ${\cal O} \equiv {\mathcal{RT}}$ (${\cal O} \equiv {\cal R}$), with $\cal R$ being a spatial operation. In Table~\ref{Table:Orbital_conductivity}, we list the symmetry restrictions of all independent components of the OC for different polarization directions of OAM and transport geometries.
\begin{table}[h]
    \centering
    \renewcommand{\arraystretch}{1.4}   
    \setlength{\tabcolsep}{8 pt}  
    \caption{The symmetry restrictions of the Drude and intrinsic components of the orbital conductivity (OC). The cross (\xmark) and tick (\cmark) marks indicate that the corresponding response tensor is symmetry forbidden and allowed, respectively. For brevity, we denote the orbital Hall conductivity $\sigma_{b;c}^{L_a}$ by $abc$, where $\{a,b,c\}=\{x,y,z\}$. Here, ${\cal M}_{a}$ and ${\cal C}_{n}^a$ represent mirror and $n$-fold rotation symmetry operations along the $a$ direction.}
    \begin{tabular}{c c c c c c c c c c}
    \hline \hline
    OC & $\sigma^{L_a}_{b;c}$ & $\mathcal{C}^x_{2,4,6}\, , \mathcal{M}_x$ & $\mathcal{C}^y_{2,4,6} \, ,\mathcal{M}_y$ & $\mathcal{C}^z_{2,4,6} \, , \mathcal{M}_z$ & {$\mathcal{M}_x\mathcal{T}$} & {$\mathcal{M}_y\mathcal{T}$} & {$\mathcal{M}_z\mathcal{T}$} & $\mathcal{C}^z_4 \mathcal{T}$ \\
    \hline \hline
    
    \multirow{8}{*}{$\sigma^{L_a,\mathrm{Dr}}_{b;c}$} & $zxz$, $zzx$, $yxy$, $yyx$  & \cmark & \xmark & \xmark & \xmark & \cmark & \cmark & \xmark \\
    \cline{2-9}

    & $zyz$, $zzy$, $xxy$, $xyx$ & \xmark & \cmark & \xmark & \cmark & \xmark & \cmark & \xmark \\
    \cline{2-9}

    & $xxz$, $xzx$, $yyz$, $yzy$ & \xmark & \xmark & \cmark & \cmark & \cmark & \xmark & \cmark \\
    \cline{2-9}

    & $zxy$, $zyx$, $xyz$, & \multirow{2}{*}{\cmark} & \multirow{2}{*}{\cmark} & \multirow{2}{*}{\cmark} & \multirow{2}{*}{\xmark} & \multirow{2}{*}{\xmark} & \multirow{2}{*}{\xmark} & \multirow{2}{*}{\cmark} \\
    & $xzy$, $yxz$, $yzx$ &  &  &  &  &  &  & \\
    \cline{2-9}
    
    & $zxx$, $zyy$  & \xmark & \xmark & \cmark & \cmark & \cmark & \xmark & \cmark \\
    \cline{2-9}

    & $xxx$, $xyy$, $xzz$ & \cmark & \xmark & \xmark & \xmark & \cmark & \cmark & \xmark \\
    \cline{2-9}

    & $yyy$, $yxx$, $yzz$  & \xmark & \cmark & \xmark & \cmark & \xmark & \cmark & \xmark \\
    \cline{2-9}

    & $zzz$  & \xmark & \xmark & \cmark & \cmark & \cmark & \xmark & \xmark \\

    \hline
    \multirow{8}{*}{$\sigma^{L_a,\mathrm{OBC}}_{b;c}$} & $zxz$, $zzx$, $yxy$, $yyx$  & \cmark & \xmark & \xmark & \cmark & \xmark & \xmark & \cmark \\
    \cline{2-9}

    & $zyz$, $zzy$, $xxy$, $xyx$ & \xmark & \cmark & \xmark & \xmark & \cmark & \xmark & \cmark \\
    \cline{2-9}

    & $xxz$, $xzx$, $yyz$, $yzy$ & \xmark & \xmark & \cmark & \xmark & \xmark & \cmark & \xmark \\
    \cline{2-9}

    & $zxy$, $zyx$, $xyz$, & \multirow{2}{*}{\cmark} & \multirow{2}{*}{\cmark} & \multirow{2}{*}{\cmark} & \multirow{2}{*}{\cmark} & \multirow{2}{*}{\cmark} & \multirow{2}{*}{\cmark} & \multirow{2}{*}{\xmark}  \\
    & $xzy$, $yxz$, $yzx$ &  &  &  &  &  &  & \\
    \cline{2-9}

    & $zxx$, $zyy$  & \xmark & \xmark & \cmark & \xmark & \xmark & \cmark & \xmark \\
    \cline{2-9}

     & $xxx$, $xyy$, $xzz$ & \cmark & \xmark & \xmark & \cmark & \xmark & \xmark & \xmark \\
    \cline{2-9}

    & $yyy$, $yxx$, $yzz$  & \xmark & \cmark & \xmark & \xmark & \cmark & \xmark & \xmark  \\
    \cline{2-9}
    
    & $zzz$  & \xmark & \xmark & \cmark & \xmark & \xmark & \cmark & \cmark  \\
    
    \hline

    \end{tabular}
    \label{Table:Orbital_conductivity}
\end{table}



\newpage


\FloatBarrier
\section{Magnetic point group symmetry of orbital-splitter current} \label{Appendix_Sec:MPG_OSC}


In this section, we carry out a comprehensive symmetry analysis of the $122$ magnetic point groups (MPGs) to identify those that allow a finite OSC. For each MPG, we examine whether the transverse orbital current is allowed while both the transverse charge current and the longitudinal orbital current are forbidden. Accordingly, we summarize the symmetry-allowed MPGs in Table~\ref{Table:OSC} for the six independent OSC components with out-of-plane orbital polarization in the $xy$, $yz$, and $zx$ transport planes.

\newcommand{\hy}{\mbox{-}}

\begin{table}[h]
\centering
\renewcommand{\arraystretch}{1.8}
\setlength{\tabcolsep}{15 pt}
\caption{Magnetic point groups that allow a finite orbital-splitter current for the indicated tensor components.}
\begin{tabular}{ c c}
\hline\hline
$\sigma^{L_a}_{b;c}$ &  MPGs \\ 
\hline\hline 

\multirow{6}{*}{\centering $\sigma^{L_z}_{y;x}$, $\sigma^{L_z}_{x;y}$}  & $2.1, 2.1', m.1, m.1', 2/m.1, 2/m.1', 2'/m, 2/m', 222.1, 222.1', mm2.1, mm2.1', m'm2', mmm.1, mmm.1' $ \\

& $m'mm, m'm'm', 422.1, 422.1', 4'22', 4mmm.1, 4mmm.1', -42m.1, -42m.1', -4'2m', 4.mmm.1$ \\

& $ 4/mmm.1', 4/m'mm, 4'/mm'm, 4'/m'm'm, 4/m'm'm', 32.1, 32.1', 3m.1, 3m.1', -3m.1, 3m.1', -3'm $ \\

& $ -3'm', - 622.1, 622.1', 6'22', 6mm.1, 6mm.1', 6'mm', -6m2.1, -6m2.1', -6'm'2, -6'm2', 6/mmm.1$ \\

& $ 6/mmm.1', 6/m'mm, 6'/mmm', 6'/m'mm', 6/mm'm', 6/m'm'm', 23.1, 23.1',  m\hy3m.1, m\hy3.1', m'\hy3' $ \\

& $ 432.1, 432.1', 4'32', -43m.1, -43m.1', -4'3m', m\hy3m.1, m\hy3m.1', m'\hy3'm, m\hy3m', m'\hy3'm'$ \\

\hline

\multirow{6}{*}{\centering $\sigma^{L_x}_{y;z}$, $\sigma^{L_x}_{z;y}$} & $2.1, 2.1', m.1, m.1', 2/m.1, 2/m.1', 2'/m, 2/m', 222.1, 222.1', 2'2'2, mm2.1, mm2.1', m'm2', m'm'2 $ \\

& $ mmm.1, mmm.1', m'mm, mm'm', m'm'm', 4.1, 4.1', 4',-4, -4.1', -4', 4/m, 4/m.1', 4/m', 4'/m, 4'/m' $ \\

& $ 422.1, 422.1', 4'22', 42'2', 4mmm.1, 4mmm.1', 4m'm', -42m.1, -42m.1', -4'2m', -42'm', 4.mmm.1$ \\

& $ 4/mmm.1', 4/mm'm', 4/m'mm, 4'/mm'm, 4'/m'm'm, 4/m'm'm', 3m.1', 6, 6.1', -6, -6.1', 6/m$ \\

& $ 6/m.1', 6/m', 6'/m, - 622.1, 622.1', 62'2', 6mm.1, 6mm.1', 6m'm', -6m2.1, -6m2.1', -6m'2', -6'm'2 $ \\

& $ 6/mmm.1, 6/mmm.1', 6/mm'm', 6/m'mm, 6'/mmm', 6/mm'm', 6/m'm'm', 23.1, 23.1',  m\hy3m.1 $ \\

& $ m\hy3.1', m'\hy3', 432.1, 432.1', 4'32', -43m.1, -43m.1', -4'3m', m\hy3m.1, m\hy3m.1', m'\hy3'm, m\hy3m', m'\hy3'm'$ \\

\hline

\multirow{6}{*}{\centering $\sigma^{L_y}_{z;x}$, $\sigma^{L_y}_{x;z}$} & $ 222.1, 222.1', 2'2'2, mm2.1, mm2.1', m'm'2, mmm.1, mmm.1', m'mm, mm'm', m'm'm' $ \\

& $  4.1, 4.1', 4',-4, -4.1', -4', 4/m, 4/m.1', 4/m', 4'/m, 4'/m', 422.1, 422.1', 4'22'$ \\

& $ 42'2', 4mmm.1, 4mmm.1', 4m'm', -42m.1, -42m.1', -4'2m', -42'm', 4.mmm.1, 4/mmm.1', 4/mm'm'$ \\

& $ 4/m'mm, 4'/mm'm, 4'/m'm'm, 4/m'm'm', 3m.1', 6, 6.1',  -6, -6.1', 6/m, 6/m.1', 6/m', 6'/m, - 622.1 $ \\

& $  622.1', 6'22', 62'2', 6mm.1, 6mm.1', 6'mm', 6m'm', -6m2.1, -6m2.1', -6m'2', -6'm2', 6/mmm.1 $ \\

& $ 6/mmm.1', 6/mm'm', 6/m'mm, 6'/mmm', 6'/m'mm', 6/mm'm', 6/m'm'm', 23.1, 23.1',  m\hy3m.1, m\hy3.1'$ \\

& $ m'\hy3', 432.1, 432.1', 4'32', -43m.1, -43m.1', -4'3m', m\hy3m.1, m\hy3m.1', m'\hy3'm, m\hy3m', m'\hy3'm'$ \\

\hline
\hline

\end{tabular}

\label{Table:OSC}
\end{table}

\newpage


\section{Orbital and charge current in \ch{FeSb2} in the absence of spin--orbit coupling} \label{Appendix_Sec:Orbital_charge_current_NoSOC}


\begin{figure}[h]
    \centering
    \includegraphics[width=1\linewidth]{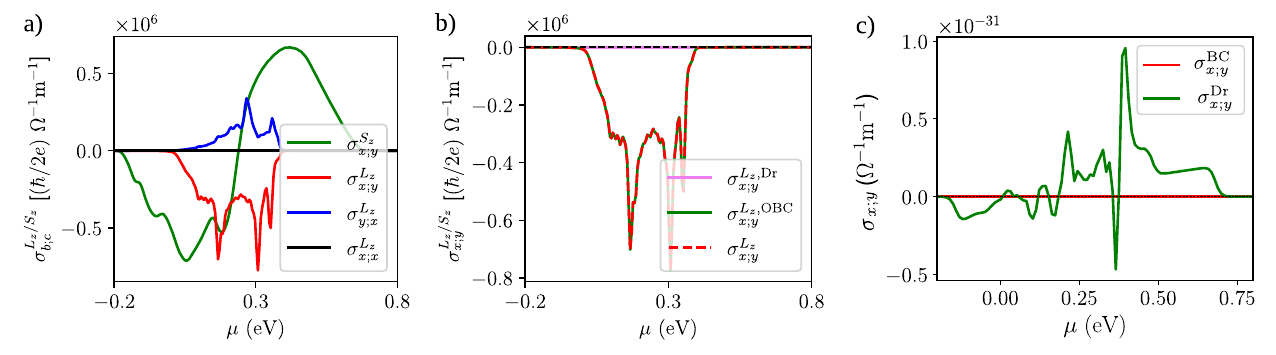}
    \caption{
    (a) Spin-splitter current and three components of the orbital current in \ch{FeSb2}. In the absence of spin--orbit coupling, the transverse orbital currents are comparable to the spin-splitter current, whereas the longitudinal component $\sigma^{L_z}_{x;x}$ vanishes. (b) Due to mirror symmetries, the Drude contribution to $\sigma^{L_z}_{x;y}$ is absent, and the response is entirely governed by the orbital Berry curvature. The same behavior holds for $\sigma^{L_z}_{y;x}$. (c) The charge current from both Drude and Berry-curvature contributions also vanishes. This is consistent with the above-discussed symmetry restrictions.
    }
    \label{fig_5_OSC_nosoc}
\end{figure}



\section{Linear charge current in \ch{FeSb2} with spin--orbit coupling} \label{Appendix_Sec:Charge_current_FeSb2}

\begin{figure}[h]
    \centering
    \includegraphics[width=0.4\linewidth]{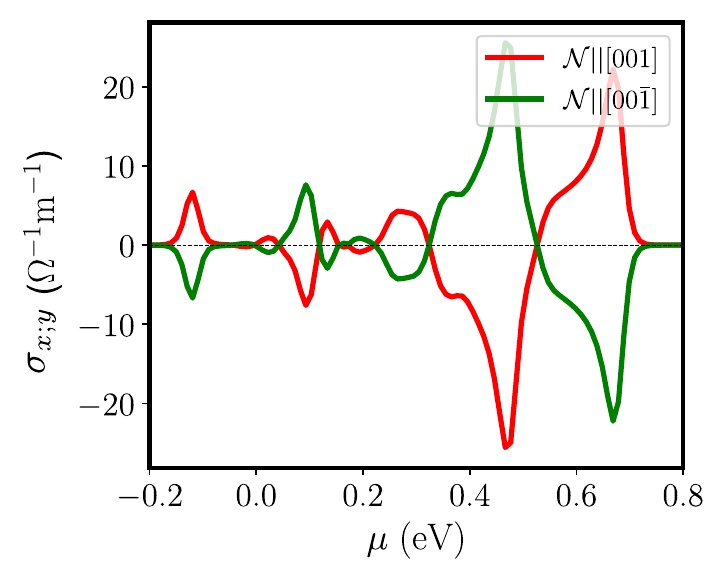}
    \caption{Anomalous Hall current (AHC) in \ch{FeSb2} in the presence of finite SOC. Red and green curves denote the AHC for two opposite orientations of the N\'eel vector along the $\hat{z}$ axis. The Hall signal is negligible compared with the transverse spin and orbital current components.}
    \label{fig_6_ahe}
\end{figure}

With finite SOC, altermagnets can exhibit an anomalous Hall effect (AHE)~\cite{Gonzalez_prl23, Attias_prb2024, Reichlova_NC2024, Sato_prl2024, Tschirner_APL2023}. Assuming $\lambda_{x0}=\lambda_{y0} = 10~\mathrm{meV}$ and $\lambda_{z0} = 5~\mathrm{meV}$, we calculate the AHE conductivity as a function of chemical potential for opposite orientations of the N\'eel vector $\mathcal{N}$, as shown in Fig.~\ref{fig_6_ahe}. The AHE reaches values of order $20~\Omega^{-1}\mathrm{m}^{-1}$ and reverses sign when the N\'eel vector is reversed, confirming its Berry-curvature origin and its $\cal T$-odd character. The charge conductivity is at least four orders of magnitude smaller than the transverse angular-momentum conductivities (see Fig.~\ref{Fig1_FeSb2}), so the accompanying transverse charge current remains negligible in practice, $(2e/\hbar)\sigma^{L_a/S_a}_{x;y}\gg\sigma_{x;y}$. Thus, \ch{FeSb2} still generates an almost pure transverse angular-momentum current.


\newpage


\section{Magnetization switching driven by SSC, OSC, and their combined effect} \label{Appendix_Sec:M_switching}



\begin{figure}[h]
    \centering
    \includegraphics[width=1\linewidth]{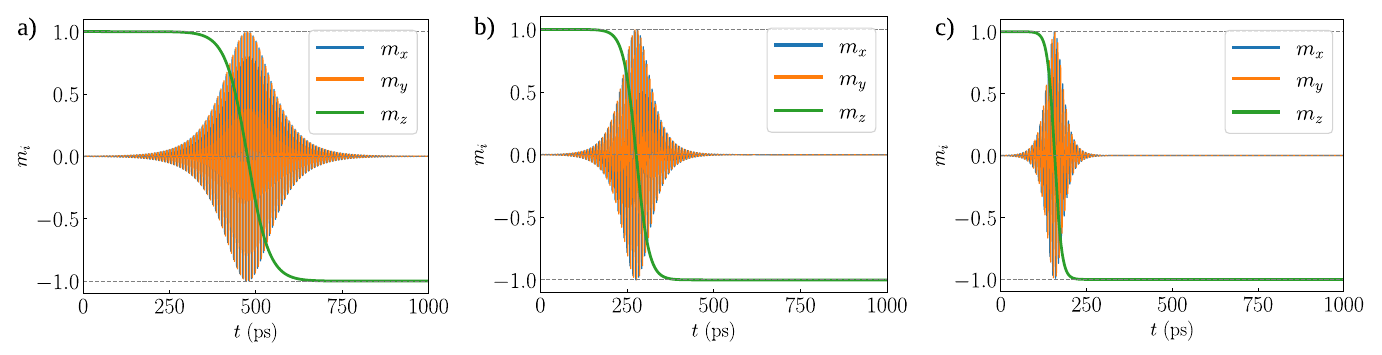}
    \caption{The switching time decreases progressively when considering (a) only the spin-splitter current, (b) only the orbital-splitter current, and (c) their combined contribution within the Landau--Lifshitz--Gilbert equation. The inclusion of the damping-like orbital torque reduces the ferromagnetic switching time by nearly a factor of three compared with the case when only spln-splitter current is applied.}
    \label{fig_7_switching}
\end{figure}


\twocolumngrid

\bibliography{Ref}
\end{document}